# Effect of sodium pyrophosphate and understanding microstructure of aqueous LAPONITE® dispersion using dissolution study

Khushboo Suman, Mohit Mittal and Yogesh M Joshi*

Department of Chemical Engineering, Indian Institute of Technology Kanpur 208016, India

*Corresponding author, E-Mail: joshi@iitk.ac.in

**Abstract**

In this work, we investigate physical origin of ergodicity breaking in an aqueous colloidal dispersion of synthetic hectorite clay, LAPONITE®, at pH 10 by performing dissolution and rheological experiments with monovalent salt and tetrasodium pyrophosphate solution. We also study the effect of interface, nitrogen and paraffin oil on the same. Dissolution experiments carried out for dispersions with both the interfaces show similar results. However, for samples with nitrogen interface, all the effects are observed to get expedited in time compared to paraffin oil interface. When kept in contact with water, 1.5 wt. % and 2.8 wt. % colloidal dispersion at pH 10 swell at small ages, while do not swell at large ages. The solution of tetrasodium pyrophosphate, interestingly, dissolves the entire colloidal dispersion samples with pH 10 irrespective of the concentration of clay. Experiments carried out on colloidal dispersions prepared in water having pH 13 demonstrate no effect of water as well as sodium pyrophosphate solution on the same suggesting a possibility of the presence of negative charge on edge at that pH. We believe that all the behaviors observed for samples at pH 10 can be explained by an attractive gel microstructure formed by edge-to-face contact. Furthermore, the absence of swelling in old colloidal dispersion at pH 10 and dissolution of the same by sodium pyrophosphate solution cannot be explained by merely repulsive interactions. This behavior suggests that attractive interactions originating from edge–to–face contact play an important role in causing ergodicity breaking in the colloidal dispersions at pH 10 at all the ages irrespective of the clay concentration. We further substantiate the presence of fractal network structure formed by interparticle edge-face association using rheological tools and cryo-TEM imaging. We also conduct a comprehensive study of the effect of tetrasodium pyrophosphate in the sol-gel transition of LAPONITE® dispersion.

## I. Introduction:

In colloidal systems, the microstructure of the same is one of the most important aspects that determines the physical behavior (1, 2). The microstructure is expected to become more intricate with enhanced complexities of the building blocks or elements that constitute the same (3). In addition, at a greater concentration of colloidal particles, an increase in viscosity strongly hinders the translational diffusivity rendering limited access to the phase space causing ergodicity breaking (3-5). All these aspects pose strong challenges to the material scientists, physicists, chemists and engineers in designing these materials for useful end use. In this paper, we investigate some of the facets of the microstructure of an aqueous colloidal dispersion of synthetic hectorite clay, LAPONITE® XLG whose building blocks are composed of anisotropically shaped particles with uneven charge distribution. It has been observed that beyond a certain concentration, the aqueous dispersion of this clay spontaneously transforms itself from a free-flowing liquid to a soft solid that supports its own weight (6-8). A lot is said in the literature about the microstructure of aqueous dispersion of this clay that renders it a soft solid-like state, which is also thermodynamically out of equilibrium (8-12). Many groups have invented and reinvented the phase diagram for this system with little agreement with each other. The point of contention in the prevailing literature was whether aqueous dispersion of LAPONITE® above 2 wt. % forms an attractive gel or a repulsive glass. However, very recently Suman and Joshi (13) critically analyzed the various developments in understanding the phase behavior of the colloidal dispersion of this clay and proposed a mechanism for microstructural evolution. They also proposed a state diagram wherein the nonergodic state has been unequivocally identified as an attractive gel state irrespective of the concentration of the clay. Although the gel state has been verified using various characterization techniques such as microscopy, rheology, and electrochemical analysis, the results of the dissolution (dilution) experiments have been interpreted in the literature in favor of both the gel and the glass state. Furthermore, there has been comparatively less work with regard to effect of multivalent salt such as tetrasodium pyrophosphate on the microstructure of aqueous LAPONITE® XLG dispersion. In this work, we address these issues that lead to new insights into this subject.

The microstructure of aqueous LAPONITE® dispersion has been investigated by a variety of characterization techniques such as static and dynamic scattering (14, 15), bulk rheology (16-19), microrheology (20, 21), microscopy (13, 19), electrochemical analysis (18, 22), simulations (23-25), etc. At low concentration of LAPONITE® (above 1 wt. % but below 2 wt. %) there is a consensus that the clay particles form a space-filling network through positive edge - negative face attractive bond (9, 10, 26-28). At high concentration of the clay (above 2 wt. %), however, there is a disagreement among various groups regarding the nature of microstructure. Many groups propose that ergodicity breaking in high concentration colloidal dispersion is also due to a space-filling network produced by the positive edge-negative surface attractive bond (10, 24, 29-32). Contrary to a proposal of attractive gel, various groups claim that above 2 wt. % the dispersion forms a repulsive or Wigner glass, wherein the clay particles are self-suspended in the repulsive environment without physically touching each other (9, 15, 28). One of the prominent experimental works that supported the repulsive glass proposal carried out small angle X-Ray scattering studies and established their inference on the position of the peak in the structure factor when plotted against the wave vector. However, Suman and Joshi (13) analyzed the SAXS results and, based on the findings of Greene *et al.* (33), observed that decoupling of intensity into a product of form factor and structure factor is a poor approximation for a disk-like particle having aspect ratio around 25, which also possess dissimilar charges. The very fact that Greene *et al.* (33) suggest that, owing to high aspect ratio and high concentration of the particles, decoupling approximation leads to spurious peaks in structure factor, Suman and Joshi (13) concluded that SAXS data cannot be interpreted at its face value to infer repulsive glassy state. On the other hand, Jatav and Joshi (19) performed rheological experiments on a large number of dispersions with various concentration of LAPONITE® and salt and observe that the various rheological characteristic features of the same as it spontaneously transforms from a sol to a soft solid like state are identical in every aspect to the sol–gel transition in polymeric systems undergoing a chemical crosslinking reaction.

While various more topical studies unequivocally show the presence of an attractive gel state in the aqueous dispersion of LAPONITE®, results of some of the dissolution studies have been interpreted otherwise. In a recent study (34) LAPONITE® dispersions of various concentrations were kept in contact with deionized water to discern the origin of ergodicity breaking. For 1.5 wt. % dispersion, the dissolution experiment did not show any change in the solid-like state of the same. On the other hand, dissolution experiment with 3 wt. % dispersion, right after the dynamic arrest, led to melting of the same. Interestingly, the dissolution experiment performed one week after the dynamic arrest in 3 wt. % dispersion, however, showed swelling of the dispersion. It was claimed that 1.5 wt. % dispersion, due to the presence of clay bonds (positive edge to negative face), does not melt or undergo swelling (34). Conversely, it was proposed that the high concentration dispersion melts in the presence of water due to the presence of repulsive interactions. It was further claimed that the absence of melting in one week old 3 wt. % dispersion suggests the presence of attractive interactions in addition to the repulsive interactions (34). However, Mongondry *et al.* (10) argued that the interparticle bonds are weak and can easily be broken on the application of small stress. Furthermore, the bonds associated with edge-face interactions and van der Waals interactions, which constitute the network formation are not irreversible. This suggests that even the gel state can be reversed by the addition of water. Therefore, the inferences of the dissolution study, on one hand, has been not universally established. This aspect was very recently highlighted by Suman and Joshi (13) where they acknowledge the necessity to carefully revisit the dissolution studies so that a unified picture regarding the nature of the nonergodic state in the aqueous dispersion LAPONITE® can be established through all the characterization techniques.

The introduction of salt in clay dispersion greatly influences the aggregation dynamics. This has resulted in numerous studies exploring salt-clay interactions (35-37). Importantly, an observation that addition of monovalent salt such as NaCl in LAPONITE® dispersion, which reduces repulsion among the particles, actually accelerates the process of 'solidification' is itself counterintuitive to dominant microstructure being repulsion dominated. This phenomenon collaborates with the microstructure of LAPONITE® dispersion being in attractive gel state irrespective of the concentration. Unlike the monovalent salt such as NaCl, which expediates gelation (19, 26, 30, 38, 39), the addition of multivalent salts such as sodium pyrophosphate ($Na_4P_2O_7$) (10, 24, 40, 41) or sodium hexametaphosphate ($(NaPO_3)_6$) (42) and hydrophilic polymer (43-45) is observed to delay the network formation (13). Dissociation of $Na_4P_2O_7$ produces tetravalent anion ($P_2O_7^{-4}$), which is proposed to strongly bind to the positively charged edge causing attenuation of attraction between the edge and the face (13). On a similar front, the addition of poly(ethylene oxide) results in adsorption of the PEO chains on the LAPONITE® particles, resulting in a reduction of edge-face interaction. Therefore, the slowing down of the aging dynamics of LAPONITE® dispersion in presence of multivalent salts and hydrophilic polymer can also be used as a test for the presence of an attractive gel state (10, 40). While LAPONITE®-PEO system has widely been explored in the literature using DLS (43, 45-47), SANS (40, 48, 49) and rheology (43, 45, 46, 50), reports on LAPONITE®-$Na_4P_2O_7$ systems are few in number (40, 51). According to a review, a systematic study of the effect of $Na_4P_2O_7$ on low and high concentration of LAPONITE® dispersion is necessary in order to distinguish the contribution of repulsive and attractive interactions (9).

The sol-gel transition in LAPONITE® dispersion studied by different groups has different protocols of sample preparation. Any minute change in the sample preparation protocol can have a noticeable effect on the process of the sol-gel transition (13). Cummins (52) investigated the effect of pH adjustment in LAPONITE® dispersion using photon correlation spectroscopy. He observed that the dispersion prepared without pH modification showed faster aggregation compared to the one whose pH was adjusted to 10. Furthermore, the edge charge of a LAPONITE® particle is greatly dependent on the pH of the medium and reducing the initial pH of the dispersion makes the edges more electropositive (13). As a result of higher electropositive edges, the aggregation rate gets enhanced due to faster edge-to-face associations. Therefore, the pH of the dispersion has a pronounced effect on the sol-gel transition process. In a recent study, Shahin *et al.* (53) studied optical birefringence in LAPONITE® dispersion and observed a strong enhancement in the orientational order near the air-LAPONITE® dispersion interface which extended over length-scales beyond five orders of magnitude larger than the particle length-scale. Interestingly, such long-range orientational order was absent at the interface when the LAPONITE® dispersion was covered with paraffin oil. This clearly suggests that the nature of the interface greatly affects the sol-gel transition in LAPONITE® dispersion.

In this work, we carry out extensive dissolution experiments on a colloidal dispersion of LAPONITE® having a low and high concentration. We also investigate the effect of pH on dissolution experiments. Furthermore, we perform the dissolution experiments with sodium pyrophosphate and sodium chloride solutions and also examine the effect of interface, nitrogen and paraffin oil, on the dissolution phenomena. We also characterize the microstructure of LAPONITE® dispersion using bulk rheology and perform a systematic rheological study on the effect of tetrasodium pyrophosphate on the sol-gel transition. Lastly, we also obtain a cryo-TEM image of a high concentration LAPONITE® dispersion so as to directly relate a microstructure to the rheological findings.

## II. Materials and Experimental Methods

**Materials and Sample preparation:** LAPONITE® XLG (a registered trademark of BYK Additives) is a 2:1 smectite hectorite clay. The chemical formula of its unit crystal is given by: $Na_{+0.7}[(Si_8Mg_{5.5}Li_{0.3})O_{20}(OH)_4]_{-0.7}$ (13). A particle of LAPONITE® has a disc-like shape with a diameter in the range of 25-30 nm and a thickness of about 1 nm (13). An idealized unit cell of the hectorite clay crystal is illustrated in Fig. 1 (54, 55). It can be seen that in a clay particle, an octahedral layer of magnesia is sandwiched between two tetrahedral layers of silica. In a unit cell, six magnesium atoms of the octahedral layer and eight silicon atoms belonging to two tetrahedral layers on either side, share eight oxygen atoms. Remaining twelve oxygen atoms are bonded to eight silica atoms and are on the outer sides of the unit cell. Four hydroxyl groups, according to this idealized cell structure, have bonds with only magnesium atoms (according to van Olphen (55), in 2:1 smectite clays, hydroxyl groups form bonds only with the metal atoms belonging to octahedral layer, which in the present case is magnesium). Isomorphic substitution of divalent magnesium by monovalent lithium creates a deficiency of positive charge within the cell (55). This renders a net negative charge to the face (charge deficiency of 0.7 per unit cell) of a disc, which is balanced by positively charged sodium ions that reside in the interlayer gallery. In a LAPONITE® particle, unit cell shown in Fig. 1 repeats itself in two directions such that a single disc having 25 nm diameter contains around 1100 unit cells with the net charge on the same to be around 770 $e$ (13). In aqueous medium, sodium ions undergo dissociation giving a permanent negative charge to the faces of a clay particle. At the edge of a disc, tetrahedral silica and octahedral magnesia structures are disrupted due to breakage of primary bonds (55). As shown in Fig. 1, edge contains oxides of silicon and lithium (in a very small amount) and oxides and hydroxides of magnesium. The charge associated with the edge of a disc depends on the pH of the medium. The edge charge changes from positive to negative at an isoelectric point (or a point of zero charge) associated with the (hydrous) oxides present on the same (55). Below the isoelectric point, protonation renders a net positive charge to the edge. On the other hand, above isoelectric point, deprotonation makes the edge negatively charged. As can be seen from the unit cell, the edge of a hectorite disc consists of –Mg-OH groups, which are known to have an isoelectric point around a pH of 12 (56). The region on the edge, where tetrahedral silica is broken usually carries a negative double layer (negative charge) (55) as the isoelectric point associated with silica is around 2 (56). Remarkably, the isoelectric point of the edge of hectorite clay has been reported to be around 10.5, which suggest it to be significantly skewed towards -Mg-OH. Nonetheless, in any event, the edge of hectorite clay particle must carry negative charge beyond pH of 12.5.

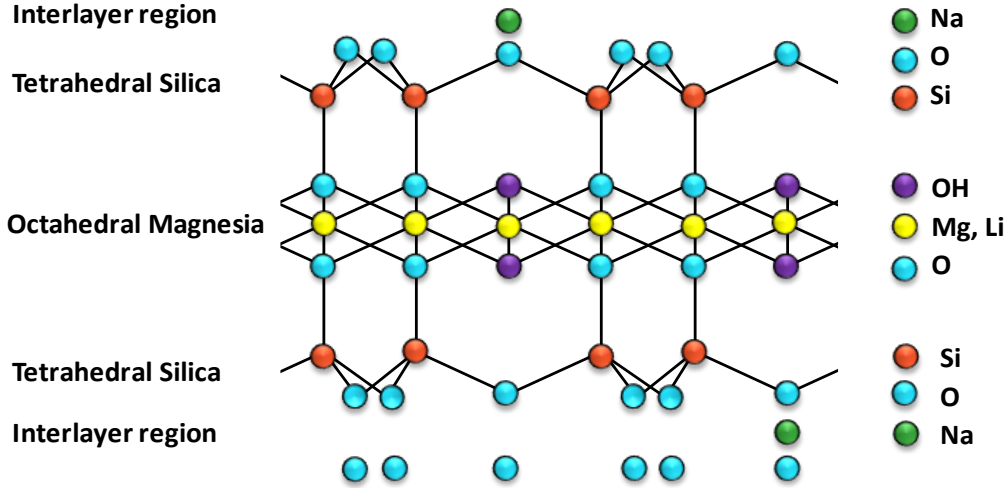

**Figure 1.** Idealized structural formula of LAPONITE® [Na$_{+0.7}$[(Si$_8$Mg$_{5.5}$Li$_{0.3}$)O$_{20}$(OH)$_4$]$_{-0.7}$] based on various proposals in the literature (54, 55).

The white powder of LAPONITE® XLG is dried at 120°C for four hours before dispersing in ultrapure Millipore water (resistivity 18.2 MΩ cm) maintained at an initial pH of 10 and 13. The basic pH is obtained by addition of NaOH. The dispersion is stirred for 30 minutes using IKA-Ultra Turrax drive, which leads to a transparent homogeneous dispersion. The chemicals (NaOH and NaCl) used in sample preparation are procured from Fischer Scientific and Na$_4$P$_2$O$_7$ from Loba Chemicals.

**Dissolution studies:** For the dissolution study, we explore two concentrations of the studied synthetic clay, 1.5 wt. % and 2.8 wt. %. During the sample preparation, soon after the stirring is complete, 5 ml of the salt-free dispersion is stored in cylindrical glass cuvettes. We employ two storage protocols. In the first protocol, the remaining volume of the cuvette is filled with nitrogen, while in the second protocol a layer of paraffin oil is placed over the free surface. All the cuvettes were sealed tightly and stored at 30 °C. The samples are monitored at frequent intervals to study the time taken by the same to support its own weight upon inversion. We report this time as the arrestation time $(t_a)$. After 2 and 10 days succeeding the arrestation, following solutions are added to the independent cuvettes: (a) deionized water having pH 10, (b) 40 mM aqueous NaCl solution, and (c) 10 mM aqueous sodium pyrophosphate ($Na_4P_2O_7$) solution. The concentrations of NaCl and $Na_4P_2O_7$ are chosen in a manner such that the ionic strength of sodium ions is the same in both the solutions (b and c). The stock solutions are prepared in ultrapure Millipore water having pH 10. A similar procedure was also carried out for the colloidal dispersion samples prepared with Millipore water having pH 13. (It should be noted that the LAPONITE® particle is reported to be completely stable at pH 13 (57)).

**Rheological measurements:** We prepare a low (1.5 wt. %) and high (2.8 wt. %) concentration LAPONITE® dispersion to investigate the nature of nonergodic state rheologically. Furthermore, in order to systematically examine the effect of tetrasodium pyrophosphate, we prepare 2.8 wt. % dispersion with 3 mM NaCl and varying concentrations of tetrasodium pyrophosphate in the range of 0-0.5 mM. All the rheological experiments are performed on TA Instruments, DHR 3 rheometer with concentric cylinder geometry having a cup diameter of 30 mm and a gap of 1 mm. In all the rheological experiments, freshly prepared samples are loaded into the shear cell and are subjected to cyclic frequency sweep over a range of 0.5-25 rad/s at a constant stress of 0.1 Pa. A thin layer of low viscosity silicone oil is applied over the free surface of the sample to prevent the evaporation loses during the long duration of the measurement. The rheological measurements are conducted at a constant temperature of 30 °C unless otherwise mentioned.

**Cryo-TEM imaging:** We prepare a 2.8 wt. % LAPONITE® dispersion using the above protocol and filter the sample using a Millex-HV 0.45$\mu$m Sterile filter. The filtered sample having an age of 110 hours is loaded on a 400-mesh holey carbon grid by the process of vitrification wherein the loaded grid is immediately plunged in liquid ethane using a reservoir of liquid nitrogen. The vitrified sample was subsequently transferred to the cryo holder for examining in a FEI-Tecnai G2 12 Twin TEM 120kV. We take multiple images at different defocus values to ensure that there is no loss of microstructural information.

## III. Results and Discussions:

III.1 Dissolution experiments:

Incorporation of LAPONITE® in water (for a concentration of 1.5 and 2.8 wt. %) eventually leads to an arrested state, wherein dispersion acquires enough strength to sustain its weight against gravity. The time required to achieve this state, which we define as arrestation time, is found to depend on the clay concentration, pH and the type of interface: nitrogen or paraffin oil, as shown in table 1. As expected, and as reported in the literature, dispersion with a greater concentration of LAPONITE® takes lesser time to arrest (13). However, the nature of the interface appears to have a noticeable effect on gelation time, with samples having paraffin oil interface delaying the process of gelation. The expedited structure formation in case of air interface is due to the cooperative dynamics of particles initiating near the air interface which eventually progresses into the bulk sample (53). Furthermore, the interface induced structure formation is absent in case of paraffin oil samples owing to the organophobic nature of the colloidal dispersion, consequently causing the delay in gelation. Interestingly, the pH of the aqueous medium also affects the gelation behavior, as samples with pH 13 show substantially expedited gelation. It is observed that at pH 13, the 2.8 wt. % colloidal dispersion undergoes gelation soon after stirring is stopped, while the samples with 1.5 wt. % takes around 6 days to undergo gelation. Interestingly, the nature of interface does not show any effect on the gelation time for samples with pH 13.

**Table 1:** Arrestation time $(t_a)$ for the 1.5 wt. % and 2.8 wt. % colloidal dispersion at pH 10 and 13.

|  | pH 10 | | pH 13 | |
| --- | --- | --- | --- | --- |
|  | Nitrogen interface | Oil interface | Nitrogen interface | Oil interface |
| 1.5 wt. % | 7 days | 38 days | 6 days | 6 days |
| 2.8 wt. % | 2 days | 3 days | Within an hour | Within an hour |

Next, we study the effect of the addition of water, sodium chloride and tetrasodium pyrophosphate solutions on 1.5 wt. % dispersion. Since the soft solid is fragile soon after arrestation, the dissolution studies were carried out 2 and 10 days after the arrestation. The corresponding results are reported in tables 2 (nitrogen interface) and 3 (paraffin oil interface). In 2 and 10 days old 1.5 wt. % dispersions with pH 10 and nitrogen interface, both water as well as sodium chloride solutions do not show any effect on the dispersion. However, for paraffin oil interface, the dispersion with pH 10 undergoes swelling at an age of 2 days but does not show any effect at an age of 10 days. Moreover, the addition of tetrasodium pyrophosphate solution dissolves both the types of dispersions (air and oil interface) completely irrespective of the age of dispersion.

**Table 2:** Effect of water, sodium chloride and sodium pyrophosphate solutions on an aqueous dispersion of LAPONITE® at 1.5 wt. % with nitrogen interface at pH 10 and pH 13 for experiments carried out on the mentioned number of days after arrestation.

| 1.5 wt. % with Nitrogen interface | pH 10 $t_a = 7$ days | | pH 13 $t_a = 6$ days | |
|---|---|---|---|---|
| | $t_a +2$ days | $t_a +10$ days | $t_a +2$ days | $t_a +10$ days |
| Water | No effect | No effect | No effect | No effect |
| Sodium Chloride | No effect | No effect | No effect | No effect |
| Sodium Pyrophosphate | Dissolved | Dissolved | No effect | No effect |

**Table 3:** Effect of water, sodium chloride and sodium pyrophosphate solutions on an aqueous dispersion of LAPONITE® at 1.5 wt. % with paraffin oil interface at pH 10 and pH 13 for experiments carried out on the mentioned number of days after arrestation.

| | pH 10 | pH 13 |
|---|---|---|

| 1.5 wt. % with Paraffin oil interface | $t_a = 38$ days | | $t_a = 6$ days | |
|---|---|---|---|---|
| | $t_a +2$ days | $t_a +10$ days | $t_a +2$ days | $t_a +10$ days |
| Water | Swelled | No effect | No effect | No effect |
| Sodium Chloride | No effect | No effect | No effect | No effect |
| Sodium Pyrophosphate | Dissolved | Dissolved | No effect | No effect |

As mentioned in the introduction section, there is a general consensus among various groups that aqueous dispersion of LAPONITE® having 1.5 wt. % concentration forms an attractive gel (a space-filling network structure) with an edge–to–surface bonds. We believe that a network formed by the clay particles may or may not swell when exposed to water depending upon the rigidity (or flexibility) associated with the network junctions. Therefore, there is a possibility that the network strands are flexible enough to undergo swelling in 2 days old colloidal dispersion exposed to paraffin oil interface. However, the network strands become rigid over a duration of 10 days such that no effect of water is observed. In case of nitrogen interface, the strands become rigid, particularly at the interface, in lesser duration (note that the colloidal dispersion with nitrogen interface undergoes gelation also over lesser duration due to the interface induced structure formation (53)) and therefore, no swelling is observed. Interestingly, sodium pyrophosphate solution dissolves 1.5 wt. % colloidal dispersion, while sodium chloride solution that has an equivalent concentration of $Na^+$ ions does not show any effect. This observation is in accordance with an edge–to–face interaction proposal since tetravalent pyrophosphate ions are known to get bound on positively charged edges of the hectorite clay particles obliterating edge–to–surface bond causing the dissolution of colloidal dispersion (24, 40, 41).

Tables 2 and 3 also report observations of the same experiments for 1.5 wt. % colloidal dispersions prepared in water having pH 13. It can be seen that water, sodium chloride as well as sodium pyrophosphate do not have any effect on the colloidal dispersion prepared in water having pH 13. As mentioned in the introduction, the edge of the clay particle is expected to have a negative charge at pH 13, consequently negatively charged pyrophosphate ions do not have any effect on the edge, and therefore the state of the structure. Overall the observations reported in tables 2 and 3 for 1.5 wt. % colloidal dispersion at pH 10 can be explained by a network formed by a positive edge to negative surface attractive interactions as universally accepted in the literature.

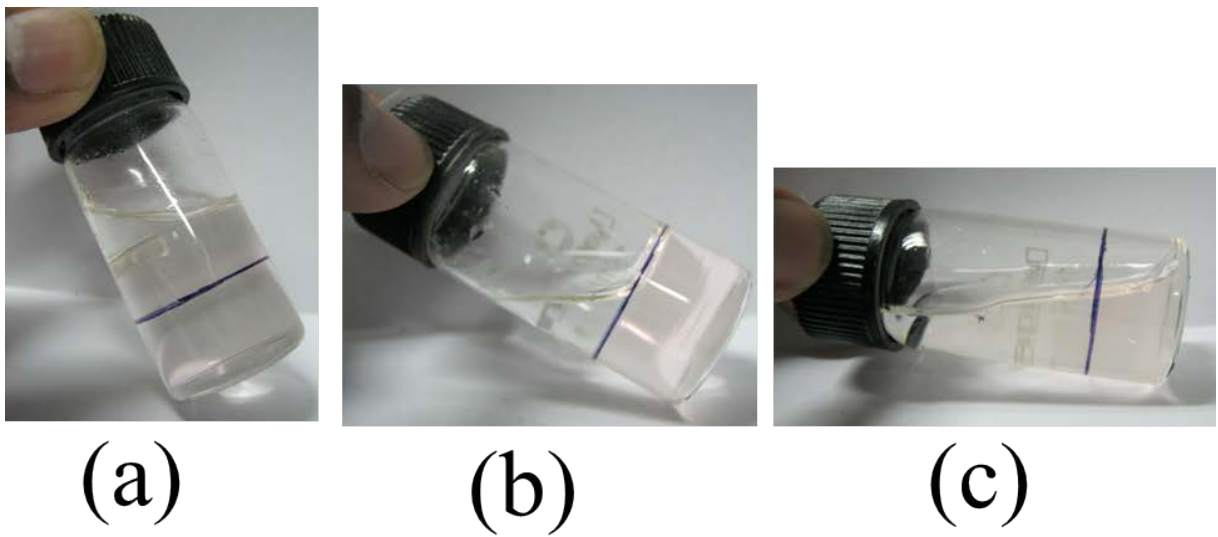

**Figure 2.** Effect of dissolution on 2.8 wt. % aqueous dispersion of LAPONITE® with paraffin oil interface at 2 days after arrestation: (a) Swelling of the arrested state on the addition of water, (b) No effect on addition of sodium chloride solution to the arrested state, and (c) Complete dissolution of the arrested state on the addition of pyrophosphate solution.

We now discuss the effect of water, NaCl and $Na_4P_2O_7$ solutions on 2.8 wt. % colloidal dispersion prepared in pH 10. As reported in table 4, dispersions with nitrogen interface having an age of $t_a + 2$ days, where $t_a = 2$ days, demonstrate swelling when put in contact with water, while swelling is not observed for $t_a + 10$ days old sample. In Fig. 2, we show the effect of addition of water, sodium chloride solution and pyrophosphate solution on 2.8 wt. % dispersion preserved with paraffin oil as interface after 2 days of arrestation and the corresponding observations are listed in Table 5. Samples with paraffin oil interface having ages of $t_a + 2$ days and $t_a + 10$ days, where $t_a = 3$ days, demonstrate swelling in the presence of water as seen in Fig. 2 (a). In order to confirm if the sample with paraffin oil interface would become rigid on an increase in time, we carried a dissolution experiment on a later day (45$^{th}$ day after arrestation). It should be noted that conducting the experiment on 45$^{th}$ day after arrestation corresponds to a total of 48 days ($t_a + 45$ days) since sample preparation. This timescale is equivalent to the dissolution experiment conducted on 1.5 wt. % dispersion on 10$^{th}$ day after arrestation since $t_a$ for 1.5 wt. % dispersion is 38 days. Interestingly, the sample having $t_a + 45$ days of age shows no effect in the presence of water, therefore signifying that the junctions have indeed become rigid with time. As reported in tables 4 and 5, for the dissolution study conducted on day 2 and 10 with pH 10 in air and paraffin oil interface, sodium chloride solution does not show any effect on 2.8 wt. % dispersion as evident in Fig. 2 (b). However, sodium pyrophosphate solution, on the other hand, is observed to dissolve the 2.8 wt. % samples at pH 10 irrespective of its interface as shown in Fig. 2 (c). For dispersion prepared with pH 13, similar to that observed for 1.5 wt. % dispersion, neither water nor sodium chloride or sodium pyrophosphate shows any effect, irrespective of the nature of the interface.

**Table 4:** Effect of water, sodium chloride and sodium pyrophosphate solutions on an aqueous dispersion of LAPONITE® at 2.8 wt. % with nitrogen interface at pH 10 and pH 13 for experiments carried out on mentioned number of days after arrestation.

| 2.8 wt. % with Nitrogen interface | pH 10 | | pH 13 | |
|---|---|---|---|---|
| | $t_a = 2$ days | | $t_a < 1$ hour | |
| | $t_a + 2$ days | $t_a + 10$ days | $t_a + 2$ days | $t_a + 10$ days |
| Water | Swelled | No effect | No effect | No effect |

| Sodium Chloride | No effect | No effect | No effect | No effect |
| Sodium Pyrophosphate | Dissolved | Dissolved | No effect | No effect |

**Table 5:** Effect of water, sodium chloride and sodium pyrophosphate solutions on an aqueous dispersion of LAPONITE® at 2.8 wt. % with paraffin oil interface at pH 10 and pH 13 for experiments carried out on mentioned number of days after arrestation.

| 2.8 wt. % with Paraffin oil interface | pH 10 $t_a = 3$ days | | | pH 13 $t_a < 1$ hour | |
|---|---|---|---|---|---|
| | $t_a +2$ days | $t_a +10$ days | $t_a +45$ days | $t_a +2$ days | $t_a +10$ days |
| Water | Swelled | Swelled | No Effect | No effect | No effect |
| Sodium Chloride | No effect | No effect | No Effect | No effect | No effect |
| Sodium Pyrophosphate | Dissolved | Dissolved | Dissolved | No effect | No effect |

The above-mentioned observations clearly suggest that all the effects that are observed for 1.5 wt. % colloidal dispersion at pH 10 is also demonstrated by 2.8 wt. % dispersion having pH 10. If microstructure of 2.8 wt. % colloidal dispersion at pH 10 is assumed to constitute a space-filling network formed by the positive edge–negative face bonds, swelling behavior of the same can be explained by a similar argument made for 1.5 wt. % dispersion at pH 10, which suggests the presence of flexible junctions that allow swelling at the beginning of the aging process. For the old samples ($t_a + 45$ days), however, junctions become rigid and do not allow swelling in water. In addition, sodium pyrophosphate solution causes the dissolution of the 2.8 wt. % dispersion having pH 10. However, no effect of sodium pyrophosphate solution on 2.8 wt. % colloidal dispersion at pH 13, further consolidates the conjecture that edges–to–face attractive bonds have a decisive role to play in pH 10, 2.8 wt. % dispersion.

Furthermore, similar to 1.5 wt. % dispersion, the effect of nitrogen and paraffin oil suggests the sol-gel process to be slow in case of paraffin oil interface system compared to that of the nitrogen interface system.

If 2.8 wt. % dispersion of LAPONITE® is assumed to possess repulsive glass-like microstructure, initial swelling of the colloidal dispersion can be understood from the repulsive interactions the clay particles share that force them to move away from each other in presence of water. However, it is difficult to comprehend no swelling of old dispersion in the presence of water, if the structure is based only on repulsive interactions. In addition, the repulsive glass scenario cannot explain the dissolution of its structure in presence sodium pyrophosphate, when sodium chloride solution having equivalent moles of $Na^+$ ions has no effect on the same. These observations conclusively rule out the possibility of the presence of a repulsive glass-like microstructure in high concentration LAPONITE® dispersion.

Experiments carried out at pH 13 for both 1.5 wt. % and 2.8 wt. % do not demonstrate any effect of sodium pyrophosphate solution. This suggests that the structure formed at pH 13 does not involve a positive edge to negative face attractive interactions. It is possible that the high concentration of $Na^+$ ions due to pH 13 causes gelation in this system via van der Waals interactions. Absence of effect of sodium pyrophosphate solution at pH 13 verifies that this gel is not composed of positive edge–to–negative face bonds and supports the conjecture that the colloidal dispersion at pH 10 forms attractive edge–to–face interactions at 1.5 as well as 2.8 wt. %. However, a systematic study of the effect of solvent pH on the gelation behavior of synthetic hectorite dispersion is the motive of our future work.

Comparison with previous Dissolution study: As discussed in the introduction, a comprehensive dissolution study has recently been conducted (34, 58) on high and low concentration clay dispersion to identify the microstructure present in the dispersion. It has been inferred that the low concentration dispersion is in the gel state since the strong interconnected clay bonds do not allow any restructuring in the sample. Furthermore, the melting of the high concentration dispersion is attributed to the formation of repulsive glass.

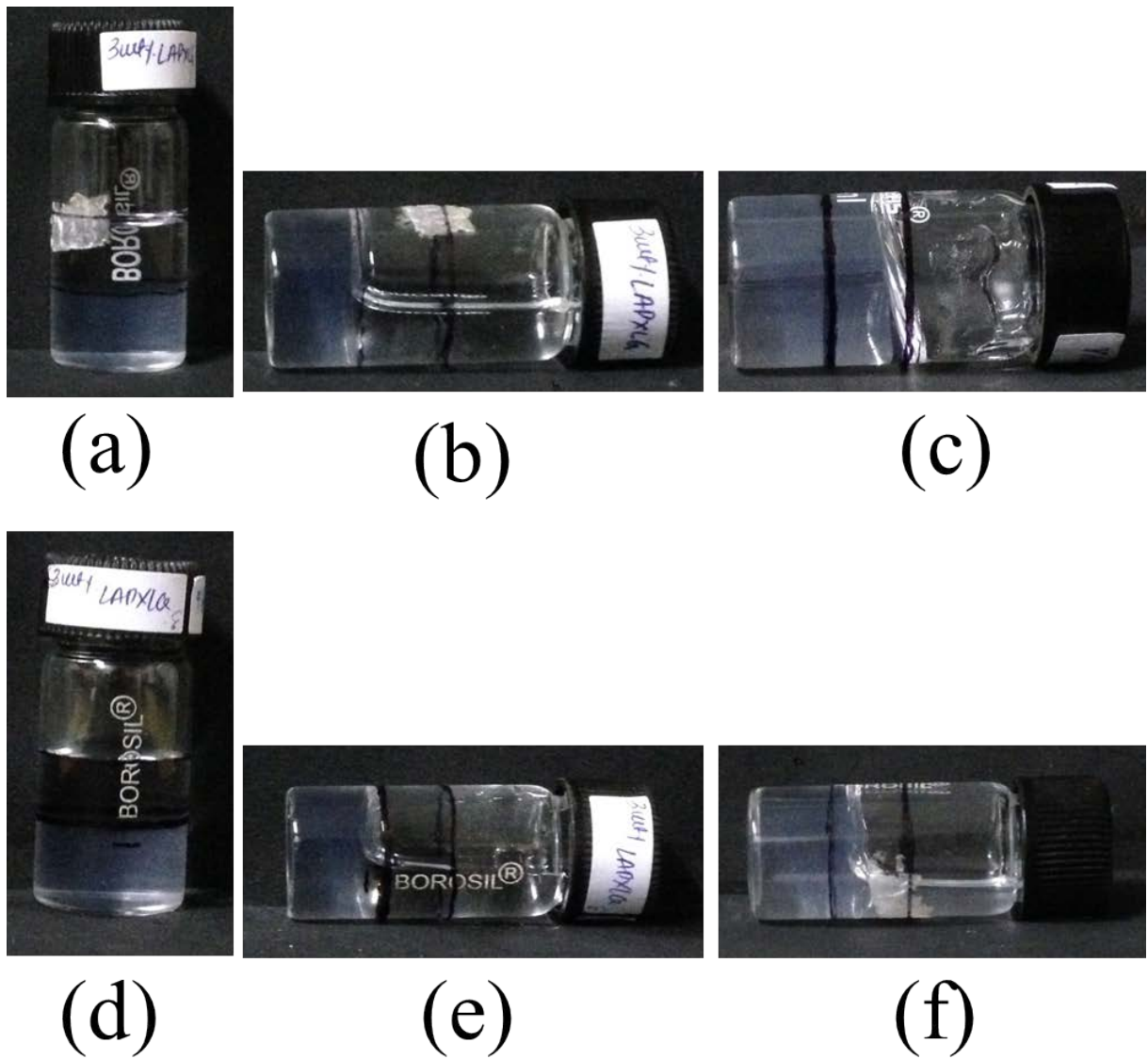

**Figure 3.** Time evolution of 3 wt. % LAPONITE® dispersion with air interface on the addition of water on $0^{th}$ day (a, b, c) and $4^{th}$ day (d, e, f) since arrestation. The initial state on the addition of water to the arrested state is shown in (a), (b), (d) and (e), while the final state wherein the swelling of the arrested state occurs, is depicted in (c) and (f).

With the motivation to understand the reported result, we prepare an identical system as used in the literature (34, 58) having 3 wt. % LAPONITE® dispersion and add water to the sample soon after arrestation. The photographic sequence of dissolution experiment is shown in Fig. 3. On addition of an equal volume of water, no effect is observed macroscopically on $0^{th}$ day sample as shown in Fig. 3 (a) and (b). Contrary to the previously reported results in the literature, with an increase in time, dissolution experiment conducted on day 0 did not lead to the melting of the semi-solid dispersion. Instead, the water gradually diffused into the network and led to the swelling of the sample a shown in Fig. 3 (c).

Furthermore, it is reported in the literature that the dissolution experiment carried before and after 3 days of arrestation leads to different behavior (58). On one hand, dissolution carried out before 3 days led to complete fluidization of the sample, while no macroscopic change was observed for dissolution carried after 3 days. The difference in observation is interpreted as the transition from repulsive glass to a disconnected house of cards structure. We perform an identical dissolution experiment on 3 wt. % clay dispersion, wherein we add water on $4^{th}$ day after arrestation. The initial state is macroscopically unaffected as seen in Fig. 3 (d) and (e). Similar to $0^{th}$ day results, eventually we observe swelling of the sample in Fig. 3 (f). Consequently, we do not observe any macroscopic difference in the results of the dissolution experiments carried on $0^{th}$ and $4^{th}$ day after arrestation which was earlier reported to be a glass-glass transition. Since the evolution of the microstructure of LAPONITE® dispersion is greatly influenced by the ambient conditions, pH, presence of impurities and sample preparation protocol (52), we believe that the difference in observation of the dissolution study can be attributed to change in any one or more of the above-mentioned parameters. Furthermore, the fact that LAPONITE® is a commercially manufactured product and may have a batch to batch variation can also contribute to the difference in the observed behavior. This suggests that a careful dissolution study does not conclusively confirm the nature of the interactions (attractive/repulsive) and the microstructure of the dispersion.

III.2 Rheological measurements:

The main inference from the dissolution experiments is that the observed behavior for no salt dispersions prepared in pH 10 water with 1.5 wt. % concentration and 2.8 wt. % concentration is qualitatively similar. The apparent difference between the two systems primarily stems from the significantly different timescales of arrestation associated with low and high concentration systems. As shown in Table 1, the typical time of arrestation for 1.5 wt. % dispersion prepared in water having pH 10 is about 7 days while 2.8 wt. % underwent arrest within 2 days in presence of nitrogen as the interface. Furthermore, there is as such no debate in the literature regarding the microstructure of low concentration dispersion. As a result, in the rheology experiments, we study the behavior of 2.8 wt. % dispersion for the signatures of gelation transition.

The rheological study gives information about the flow behavior of the viscoelastic material in terms of elastic $(G')$ and viscous $(G'')$ modulus. During the process of sol-gel transition, as the material transforms from the sol state (that possesses zero shear viscosity) to the gel state (that possesses equilibrium modulus) (59), the rheological study can serve as an important tool to describe the transition. In a seminal contribution by Winter (60) on crosslinking polymeric system undergoing a transition from a sol state to a three-dimensional crosslinked network, the system passes through a critical state with the weakest space spanning percolated network. At this point of critical gel transition, Winter and Chambon (61) observed $G'$ and $G''$ to exhibit an identical power-law dependence on frequency $(\omega)$ given by:

$$G' = G'' \cot(n\pi/2) = \frac{\pi S}{2\Gamma(n)\sin(n\pi/2)}\omega^n, \tag{1}$$

where $S$ is the gel strength of a network, $n$ is the critical relaxation exponent limited between 0 and 1, and $\Gamma(n)$ is the Euler gamma function of $n$. Consequently, the loss tangent given by $\tan\delta = \tan(n\pi/2) = G''/G'$, becomes independent of $\omega$ suggesting the sol-gel transition to be independent of the timescale of observation (60, 61). The Winter criteria (61) has been validated experimentally using bulk rheology as well as micrcorheology study. The identical power-law dependence of dynamic moduli on $\omega$ results in a continuous power-law spectrum of relaxation times given by :
$H(\tau) = \dfrac{S}{\Gamma(n)} \tau^{-n}$ (62). The power-law rheology exhibited by the critical gel state arises from the fractal nature of the percolated network. In an important contribution by Muthukumar (63)], the fractal dimension $(f_d)$ of the hierarchical network with a complete screening of the excluded volume effects can be computed with the knowledge of $n$ using the relationship:

$$f_d = 5(2n-3)/2(n-3). \tag{2}$$

Very interestingly, the fractal dimension computed using the relation proposed by Muthukumar has been observed to agree well with that obtained by scattering studies for different gel-forming systems (64-66).

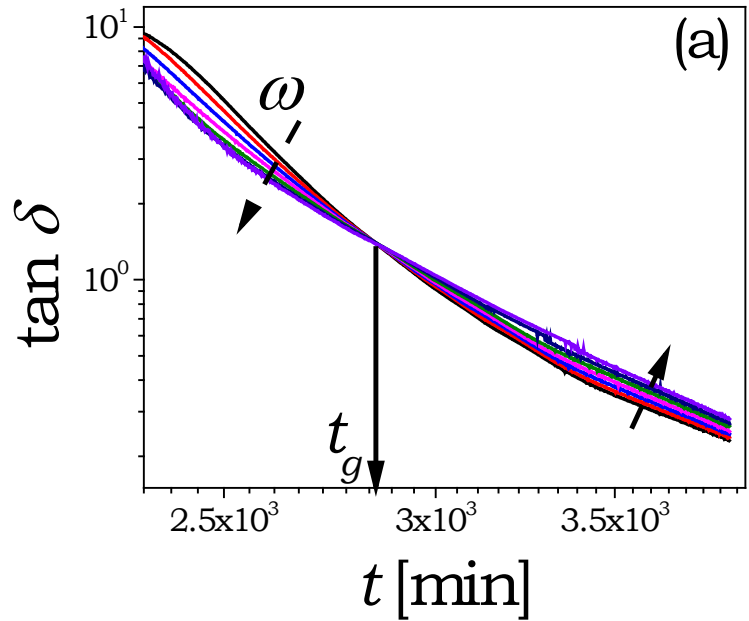

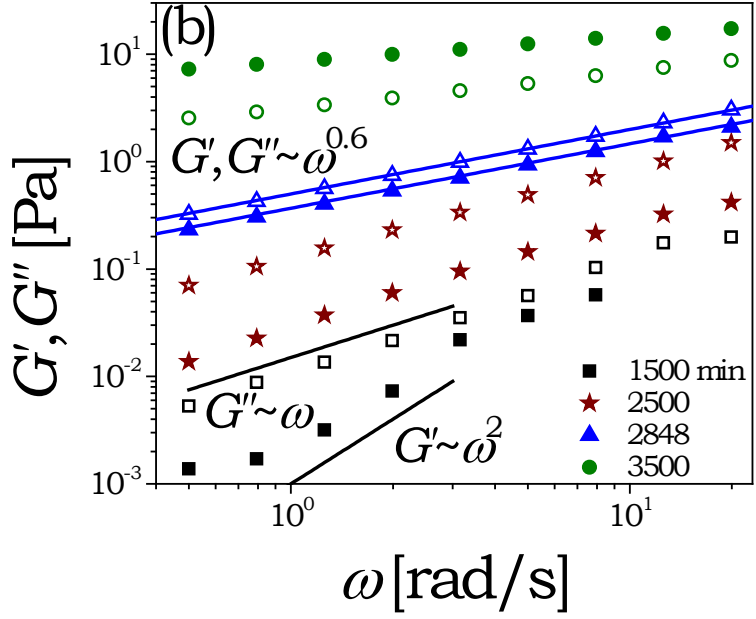

**Figure 4.** (a) Evolution of $\tan \delta$ as a function of time is plotted at different angular frequencies for 2.8 wt. % LAPONITE® dispersion. The dashed line indicates the direction of increase in the angular frequency. The arrow denotes the point of critical gelation. (b) Evolution of $G'$ and $G''$ is plotted as a function of $\omega$ at different times for 2.8 wt.% LAPONITE® dispersion. The solid line is the prediction given by Eq. (1) to the data at the critical gel state (2848 minutes).

Figure 4 (a) describes the results of a cyclic frequency sweep experiment on 2.8 wt. % LAPONITE® dispersion. The evolution of $G'$ and $G''$ is shown in Fig. 4 (b). Initially, the dispersion is liquid-like with $G'' > G'$. In addition, the terminal regime is observed at early times with $G' \sim \omega^2$ and $G'' \sim \omega$ dependence. With a gradual increase in time, the dynamic moduli grow with the rate of growth of $G'$ being greater than $G''$. Interestingly, at a certain time, both moduli exhibit identical power-law dependence on the frequency, which signifies the critical gel state wherein the system forms the weakest space spanning percolated network. The solid line in Fig. 4 (b) represents the dependence of $G'$ and $G''$ given by Eq. (1) with $S = 0.3$ Pa.s$^n$ and $n = 0.6$. Very interestingly, the identical power-law dependence has been observed in LAPONITE® dispersion using micrcorheology study as well (20, 67). Beyond the critical gel time, the frequency dependence of moduli weakens. The corresponding evolution of $\tan\delta$ for high concentration dispersion has been plotted as a function of time in Fig. 4 (a). The scaling of moduli in the terminal region suggests $\tan\delta$ to be inversely proportional to frequency as shown by the dashed line in Fig. 4 (a). Very interestingly, the iso-frequency $\tan\delta$ curves become independent of the applied frequency at the critical point. The time associated with $\tan\delta$ becoming independent of the observation timescale is known as the critical gelation time $\left(t_g\right)$. It should be noted that $t_g$ is different than the arrestation time recorded in the dissolution experiment. At the critical gel state, the largest aggregate spans the entire space which results in the weakest fractal network. However, the arrestation time denotes the post-gel state when the sample is able to withstand its own weight. The time to achieve critical gelation is 2848 minutes (~47.5 hours) for 2.8 wt. % dispersion as indicated by arrow in Fig. 4 (a). The corresponding relaxation exponent $\left(n\right)$ associated with the critical gel state is calculated using the relation $n = 2\delta/\pi$. Remarkably, the critical relaxation exponent is the same as the power-law exponent of moduli dependence on the frequency at the point of critical gelation. This equivalence validates the presence of percolated fractal network at the critical gel state as suggested by the Winter criterion (60) for both low and high concentration colloidal dispersion. At higher times, as the dependence of $G'$ and $G''$ weakens with respect to frequency, $\tan\delta$ increases proportionally with frequency. Such dependence of $\tan\delta$ denotes the post-gel state that has finite equilibrium modulus (59).

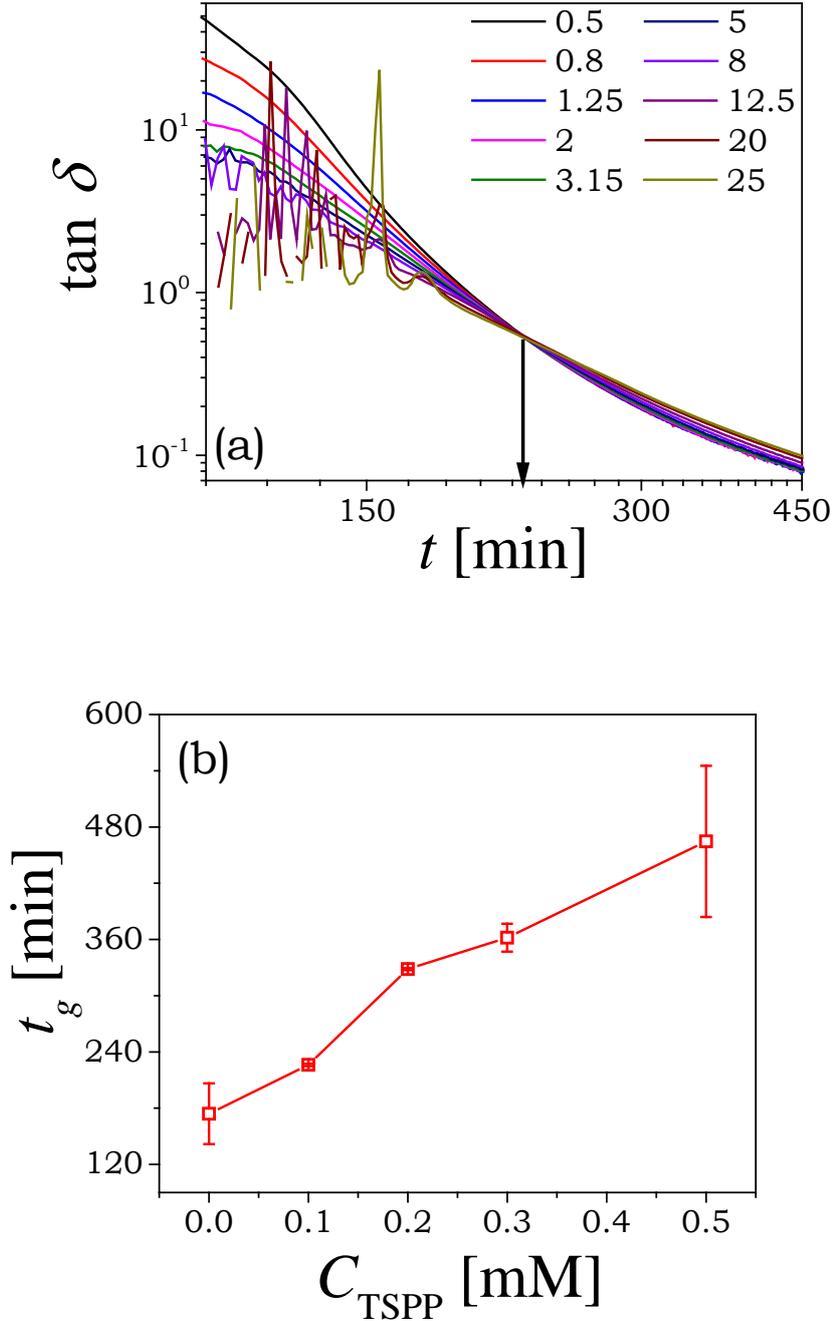

**Figure 5.** (a) Evolution of $\tan\delta$ as a function of time is plotted at different angular frequencies for 2.8 wt. % LAPONITE® dispersion with 3 mM NaCl and 0.1 mM $Na_4P_2O_7$. The arrow denotes the point of critical gelation. (b) The time to achieve critical gel transition $(t_g)$ is plotted as a function of $Na_4P_2O_7$ concentration $(C_{TSPP})$. The lines serve as a guide to the eye.

The microstructure of the colloidal dispersion is observed to be greatly dependent on the concentration of clay and the ionic strength of the salt. While the effect of monovalent salt (sodium chloride) has been widely investigated in the literature (13, 19, 39), a systematic rheological investigation of sol-gel transition in an aqueous dispersion of LAPONITE® clay on the addition of tetrasodium pyrophosphate is lacking. Therefore, next we discuss the effect of tetrasodium pyrophosphate on the rheological gelation behavior of synthetic hectorite clay dispersion. We conduct an identical rheological experiment on LAPONITE® dispersion as reported in Fig. 4 with varying concentrations of tetrasodium pyrophosphate in the range of 0-0.5 mM. Since pure 2.8 wt. % LAPONITE® dispersion takes a long time to undergo sol-gel transition (2848 minutes), and tetrasodium pyrophosphate actually prolongs it even further, it is not practically feasible in terms of timescale to perform these experiments. Consequently, we add monovalent salt (3 mM NaCl) to 2.8 wt. % LAPONITE® dispersion that screens the repulsive potential resulting in faster structure formation and reducing the gelation time by over one order of magnitude. This aids in performing the experiments over a smaller window of timeframe. The evolution of $\tan\delta$ for 2.8 wt. % colloidal dispersion having 3mM NaCl and 0.1 mM tetrasodium pyrophosphate is shown in Fig. 5 (a). It can be seen that dispersion with pyrophosphate exhibits identical qualitative behavior as shown in Fig. 4 (a) while undergoing the sol-gel transition. Furthermore, identical rheological behavior demonstrating all the characteristic features of the sol-gel transition is exhibited at all the explored concentration of pyrophosphate salt validating the Winter criterion (60). In Fig. 5 (b) we plot the evolution of the time required to achieve critical gel state as a function of tetrasodium pyrophosphate concentration $(C_{\text{TSPP}})$. The gelation time can be seen to increase with an increase in the concentration of multivalent ions in the system. The pyrophosphate system being a strong dispersant adsorbs on the edges of the positively charged clay particles. As a result, the positive charge on the edges reduces and the bond formation between the negative face and positive edge gets inhibited. Therefore, the rate of gelation gets strongly deaccelerated with the increase in ionic strength of pyrophosphate ions.

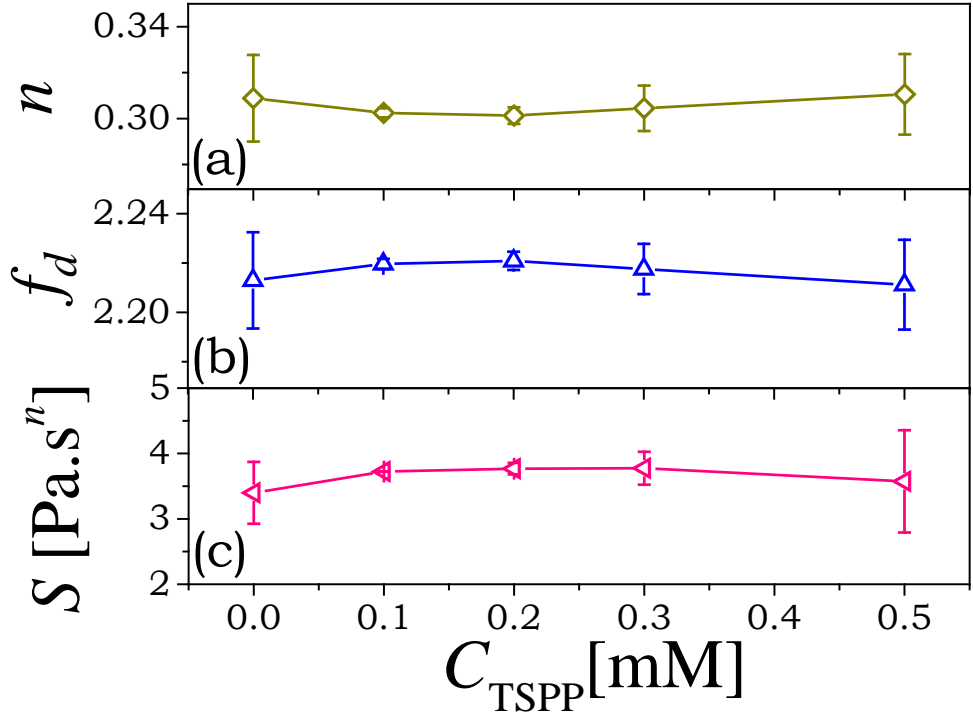

**Figure 6.** The characteristic parameters associated with the critical gelation such as (a) relaxation exponent $(n)$, (b) fractal dimension $(f_d)$ and (c) gel strength $(S)$ are plotted as a function of $Na_4P_2O_7$ concentration $(C_{TSPP})$. The lines serve as a guide to the eye.

In Fig. 6 (a) and (c), we plot the value of critical relaxation exponent $(n)$ and the gel strength $S$ obtained from the fit of Eq. (1) to the dynamic moduli data at the critical gel point as a function of pyrophosphate concentration. We also obtain the fractal dimension $(f_d)$ of the percolated network with the knowledge of $n$ using Eq. (2) (19, 68) and plot as a function of pyrophosphate concentration in Fig. 6 (b). It can be seen from Fig. 6 (a), (b) and (c) that $n$, $f_d$ and $S$ remain almost constant over the explored range of pyrophosphate concentrations, indicating the addition of pyrophosphate essentially affects the gelation kinetics keeping the fractal structure at the critical point unaffected. Such behavior could arise from screening of the positive edge charges – because of tetravalent pyrophosphate ions – thereby reducing the number of sites on the edges where the faces of other LAPONITE® particles can attach to in doing so delaying the point at which weakest percolated network (critical gel) is obtained. In a limit of sufficiently high concentration of pyrophosphate, the samples simply do not undergo gelation over the practically explorable timescales. Similar results were obtained in dissolution experiments wherein the adsorption of pyrophosphate ions on the edge of clay particle was also responsible for the dissolution of the arrested state. It must be mentioned that many sol-forming commercial variants of LAPONITE® are available in the market which contains tetrasodium pyrophosphate to inhibit the gel formation. This clearly subscribes to the fact that soft solid state formed by LAPONITE® dispersion must originate from a network with a positive edge to negative face attractive bond as the building blocks.

The discussion to this point very clearly indicates that high concentration LAPONITE® dispersion (2.8 wt. %) in the absence (Fig. 4 (a)) or presence (Fig. 5 (a)) of salts, overwhelmingly demonstrates all the rheological signatures of critical gel transition. The attractive gel state can further be confirmed by analyzing the effect of the addition of salt in LAPONITE® dispersion. In Fig. 4 (a) and Fig. 5 (b), the critical gelation time for 2.8 wt. % is observed to be 2848 minutes while for 2.8 wt. % with 3 mM NaCl and 0 mM $Na_4P_2O_7$, the time reduces to 180 minutes. It is interesting to note here that the time to achieve critical gelation decreases by over an order of magnitude on the addition of salt. This clearly suggests that the process gets expedited in time with an increase in the concentration of externally added salt. Since the introduction of salt decreases the repulsion between the clay particles, it is counter-intuitive that decreased repulsion will expedite order formation in a repulsive glass. Therefore, the possibility of the nonergodic state in high concentration clay dispersion to be a repulsive Wigner glass gets completely ruled out.

In order to have the insights about the microstructure and associate a microscopic portrait of the structure to the rheological findings, we take the cryo-TEM image of filtered 2.8 wt. % LAPONITE® dispersion. The rheological experiment suggests the time to achieve critical gel state is 2848 minutes (~ 47.5 hours). Therefore, we obtain a cryo-TEM image at a post gel state of 110 hours. The corresponding cryo-TEM image is shown in Fig. 7. The thick lines represent a LAPONITE® particle, which can be seen to be having edge-to-face bonds at different angles and overlap coin configuration (24, 25, 69), which eventually leads to a very clear percolated space spanning network as seen in Fig. 7. We believe that the cryo-TEM image is the most explicit confirmation of the presence of a gel-like microstructure formed by attractive interparticle bonds in high concentration LAPONITE® dispersion.

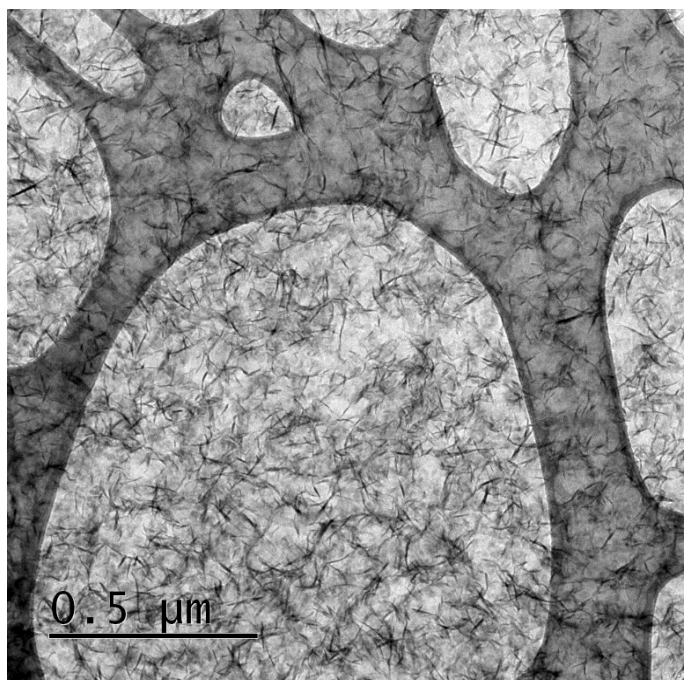

**Figure 7.** Cryo-TEM image representing the percolated network structure formed in 2.8 wt. % LAPONITE® dispersion at a post gel state (110 hours).

The observations of the present work and various other observations discussed in the introduction section very conclusively establish that the microstructure of aqueous dispersion of LAPONITE® in the higher concentration regime corresponds to the attractive gel state as also proposed by Suman and Joshi (13). Very importantly, the present work carries out the dissolution experiments, studies effect of pyrophosphate, obtains cryo-TEM images and provides further insight into the microstructure of the LAPONITE® dispersion.

## IV. Conclusion

In this work, we carry out dissolution experiments on 1.5 wt. % and 2.8 wt. % aqueous dispersion of LAPONITE® having pH 10 and 13 using deionized water, sodium chloride solution and sodium pyrophosphate solution to investigate the physical origin of ergodicity breaking, attractive or repulsive, in this system. We also perform the dissolution experiments as a function of the age of the colloidal dispersion. Furthermore, we carry out all the dissolution experiments for the colloidal dispersions having two types of interfaces: nitrogen and paraffin oil. Overall, we observe that for the experiments carried out on pH 10 system, irrespective of the clay concentration, all the effects that are shown by the nitrogen interface dispersion are also observed for the paraffin oil interface but delayed in time. This suggests that the overall evolution of microstructure is faster for nitrogen interface compared to that for paraffin oil interface.

An aqueous dispersion of LAPONITE® having 1.5 wt. % concentration at pH 10 is believed to possess a network-like structure with a positive edge-negative face attractive bonds. We observe that this system with paraffin oil interface undergoes swelling when kept in contact with water 2 days after arrestation. The aged samples with paraffin oil as well as nitrogen interface do not demonstrate swelling or any other effect in the presence of water. Interestingly, sodium chloride solution does not show any effect while sodium pyrophosphate solution always dissolves this system irrespective of the nature of interface as well as the age. For 1.5 wt. % dispersion having pH 13, neither water nor sodium chloride solution or sodium pyrophosphate solution demonstrates any effect. This result is in accordance with an attractive gel-like structure with an edge–to–face interactions postulated for low concentration (1 wt. % to 2 wt. %) colloidal dispersions having pH 10. Furthermore, no effect of sodium pyrophosphate on pH 13 system also vindicates the conjecture that proposes the edge of the clay particle to be negatively charged at this pH.

The 2.8 wt. % concentration colloidal dispersion at pH 10 and 13 also show the same behavior as that observed for 1.5 wt. % colloidal dispersion having pH 10 and 13, respectively. The time taken by the 2.8 wt. % system, however, is observed to be smaller compared to 1.5 wt. % system. Similar to the 1.5 wt. % dispersion at pH 10, all the observed behaviors of 2.8 wt. % at pH 10 can be explained by a percolated network (gel) structure wherein interactions are formed by the edge–to–face bond. On the other hand, the absence of swelling at very high ages and dissolution of dispersion when kept in contact with sodium pyrophosphate solution demonstrated by 2.8 wt. % pH 10 systems cannot be explained by merely repulsive interactions among the clay particles. The dissolution study, therefore, clearly suggests that in an aqueous dispersion of LAPONITE® at pH 10, attractive interactions originating from the positive edge–negative face contact play a decisive role in causing ergodicity breaking. Furthermore, our rheological results indicate that the high concentration sample demonstrates all the characteristic rheological signatures of the sol-gel transition, which indicates the presence of a fractal network. Additionally, we report a systematic study on the effect of tetrasodium pyrophosphate on the microstructural evolution of synthetic hectorite dispersion. The delay in time to critical gelation on increasing the tetrasodium pyrophosphate concentration further validates the presence of edge-face bonds, which are responsible for network formation. This work, therefore, on the one hand, discusses the new facets of dissolution results and address the discrepancy in the literature. On the other hand, the present work investigates the microstructure of LAPONITE® dispersion using rheology and with the help of microscopic evidence, we conclude the arrested state to be an attractive gel state.

**Acknowledgement:** Financial support from the Science and Engineering Research Board (SERB), Department of Science and Technology, Government of India is greatly acknowledged.